# Simultaneous control of optical dipole force and coherence creation by super-Gaussian femtosecond pulses in $\Lambda-$like atomic systems

**Pawan Kumar, Parvendra Kumar and Amarendra K. Sarma***

Department of Physics, Indian Institute of Technology Guwahati, Guwahati-781039, Assam, India.
*Electronic address: aksarma@iitg.ernet.in

We report a study on the optical dipole force on a beam of neutral three-level $\Lambda-$like atomic system induced by a femtosecond super-Gaussian pulse. We show that maximum coherence between the ground state |1> and the excited state |2> could be achieved using a train of femtosecond pulses. In addition, it is possible to control the trajectory of the atoms in an atomic beam by using the same scheme. The robustness of the scheme against the variation of the pulse parameters is also investigated.

**I. INTRODUCTION**

Manipulation of atoms or molecules with quantum mechanical forces has been a topic of considerable research interest for quite some time now [1]. Physicists are able to develop a broad collection of quantum mechanical tools to exert forces on atoms. It is now possible to control the positions and velocities of atoms by using optical forces so well that we can stop atoms and hold them in place for an extended time [2]. Optical force due to light, particularly laser beam, on particles such as atoms, molecules, ions etc. have been successfully exploited in as diverse areas as optical tweezers [3], atom optics [4], Bose–Einstein condensation [5], quantum information [6], etc. Many authors have studied the radiation forces exerted on neutral atoms [7-12]. Recently, owing to the recent progress in the generation of femtosecond and attosecond laser pulses, study of the mechanical effect of light on atoms and molecules is getting a tremendous boost [13-17]. In this context several studies have been reported. For example, optical force on two-level atoms by subcycle pulsed focused vector fields [13], light force on a beam of neutral two-level atoms superimposed upon a few-cycle pulsed Gaussian laser field under both resonant and off-resonant condition [14], optical dipole force on ladder-like three-level atomic systems induced by few-cycle-pulse laser fields [15], near resonant optical force [16] and trapping of nanoparticles with femtosecond pulses [3].

One important aspect of the interaction of light with atoms or molecules is the so called coherent population transfer between the quantum states of atoms and molecules [17]. Today, coherent control techniques are widely used in the fields of robust quantum dot excitation generation [18], controllable coherent population transfer in superconducting qubits [19], atomic interferometry [20, 21], high precession spectroscopy [22, 23], quantum computing [24, 25], quantum information processing [26,27], ultrafast optical switching [28-30] and population transfer in four level atoms [31]. In the last two or three decades or so, many efficient schemes such as stimulated Raman adiabatic passage (STIRAP), Raman chirped adiabatic passage (RCAP) and adiabatic rapid passage (ARP) for controlling the population transfer between the quantum states of atoms and molecules has opened new routes for controlling the various atomic and molecular processes[1,2,32-34]. In this context, another extremely important and relevant topic of interest is the creation of coherent superposition of atomic or molecular quantum states or coherence. It has potential applications in high-harmonic generation [35], electromagnetically induced transparency [36], lasing without inversion [37], four-wave mixing [38], control of chemical reactions [39,40] etc. Recently, a robust scheme is proposed for the attainment of maximum coherence with nanosecond pulses in a three level $\Lambda-$system [41]. However, it may be noted that the generation of coherence using femtosecond pulses is relatively less explored.

In this work, we discuss a scheme which enables one to control optical dipole force and coherent population transfer using a single femtosecond super-Gaussian pulse. Also, using a train of femtosecond pulses we show that in addition to creating coherence between the ground state |1> and the excited state |2> in the $\Lambda-$system, it is possible to control the trajectory of the atoms in an atomic beam. In Sec. II we present the optical Bloch equations that describe the interaction of the $\Lambda-$like three-level system with femtosecond laser pulse. Sec. III contains our simulated results and discussions followed by conclusions in Sec. IV.

## II. THE MODEL

We consider a three level $\Lambda-$like atomic system interacting with a single or a train of few-cycle-pulse laser fields. Fig.1. shows the sketch of the $\Lambda-$like three level atomic systems. The energy gap between the two lower states |1> and |2> is taken to be much less than the frequency spectrum of a single pulse in the train.

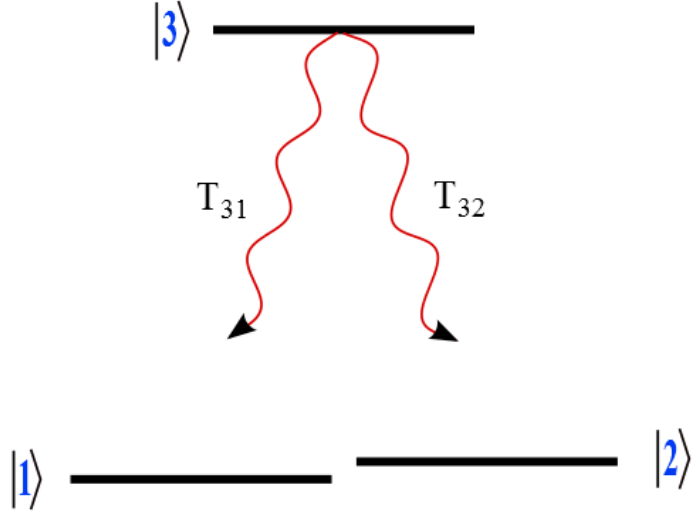

FIG.1. Sketch of the energy levels for the calculation of optical dipole force on the atoms

The electric field of a single laser pulse that interacts with the pair of energy levels |1> and |3> and |2> and |3> is linearly polarized and is given by $\widetilde{E(r,t)} = A(r,t) Cos\left((\omega_p)t - \Phi(z)\right)$, where $A(r,t)$ and $\omega_p$ are the space and time dependent field envelope and time dependent frequency of the pulse respectively. We consider a super-Gaussian pulse envelope of the general form $A(r,t) = \eta A_0 . \exp\left[-\left\{\left(\frac{|r|}{w}\right)^m + \left(\frac{|t|}{\tau_0}\right)^m\right\}\right]$, where $\eta$ denotes the polarization state of the electric field and $w$ and $\tau_0$ are the beam waist and pulse width respectively. $m$ is the order of the super-Gaussian shaped pulse. The time dependent frequency is given by $\omega_p = \omega_0 + \chi t^2 - \omega_D$, in which $\omega_0$ is the central frequency of the laser in the lab frame, $\chi$ is the nonlinear chirp rate and $\omega_D = k.v_z$ is the detuning of the transition frequency due to the translational motion, $v_z$, of the atoms. We consider that the atoms interacting with the pulse constitute an atomic beam and hence the related Doppler shift. Here $\Phi(z)$ refers to the longitudinal phase of the field. The decay and decoherence processes become significant when considering interaction with a train of pulses and thus terms accounting for these processes are added in the density matrix equations describing the temporal evolution of the system. In Eq.1 we present these equations without invoking the rotating wave approximation.

$$\frac{d\rho_{11}}{dt} = \frac{\rho_{33}}{T_{31}} + i(\Omega_{13}\rho_{31} - \Omega_{31}\rho_{13})$$

$$\frac{d\rho_{22}}{dt} = \frac{\rho_{33}}{T_{32}} + i(\Omega_{23}\rho_{32} - \Omega_{32}\rho_{23})$$

$$\frac{d\rho_{33}}{dt} = -\frac{\rho_{33}}{T_{31}} - \frac{\rho_{33}}{T_{32}} + i(\Omega_{31}\rho_{13} - \Omega_{13}\rho_{31}) + i(\Omega_{32}\rho_{23} - \Omega_{23}\rho_{32})$$

$$\frac{d\rho_{12}}{dt} = i(\omega_{21}\rho_{12} + \Omega_{13}\rho_{32} - \Omega_{32}\rho_{13})$$

$$\frac{d\rho_{13}}{dt} = -\frac{\rho_{13}}{2T_{31}} - \frac{\rho_{13}}{2T_{32}} + i(\omega_{31}\rho_{13} + \Omega_{13}(\rho_{33} - \rho_{11}) - \Omega_{23}\rho_{12})$$

$$\frac{d\rho_{23}}{dt} = -\frac{\rho_{23}}{2T_{32}} - \frac{\rho_{23}}{2T_{31}} + i(\omega_{32}\rho_{23} + \Omega_{23}(\rho_{33} - \rho_{22}) - \Omega_{13}\rho_{21}) \quad (1)$$

Here, $\Omega_{13} = \Omega_{31} = \mu_{13}E(r,t)/\hbar$ and $\Omega_{23} = \Omega_{32} = \mu_{23}E(r,t)/\hbar$ are the time dependent Rabi frequencies for the $|1\rangle$ to $|3\rangle$ and $|2\rangle$ to $|3\rangle$ electric dipole transitions respectively, with the corresponding dipole moments being $\mu_{13}$ and $\mu_{23}$. The $|2\rangle$ to $|1\rangle$ transition is dipole forbidden. $E(r,t) = \sum_{n=0}^{N-1} E(r,\widetilde{(t-nT)})$, is the electric field for a train of $N$ pulses with the pulse repetition period $T$. It is to be noted that $\rho_{ij} = \rho_{ji}^*$ represent the density matrix elements and $\omega_{ij} = \omega_i - \omega_j$, where $\omega_k$ refers to the energy level of the state $|k\rangle$. In these equations $T_{31}$ and $T_{32}$ are the spontaneous decay lifetimes for the population decay from $|3\rangle$ to $|1\rangle$ and $|3\rangle$ to $|2\rangle$ respectively. Under these conditions of excitation the expression for the optical dipole force experienced by the atoms is given by:

$$F_r = \{\mu_{13}(\rho_{13} + \rho_{31}) + \mu_{23}(\rho_{23} + \rho_{32})\} \{\sum_{n=0}^{N-1}[\nabla A(r,t-nT)]Cos\left((\omega_p(t-nT))(t-nt) - \Phi(z)\right)\} \quad (2)$$

### III. RESULTS AND DISCUSSIONS

We first present the results of treating a $\Lambda-$like atomic system with a single few cycle pulse i.e. $N = 1$. Eq. (1) and Eq. (2) are solved numerically using a standard fourth order Runge-Kutta method. We use the following typical values for the parameters in the simulation: $\omega_{31} = 1.656 \, rad/fs$, $\omega_{32} = 1.653 \, rad/fs$, $T_{31} = T_{32} = 1ns$, $\Omega_{13} \approx \Omega_{23} = \Omega = 0.37 \, rad/fs$, $\omega_0 = 1.65 \, rad/fs$, $\chi = \pm 0.0037 \, fs^{-3}$, $w = 100 \, \mu m$, $\tau_0 = 10 \, fs$ and $v_z = 100 m/s$.

We choose an off-axis representative point at $r = 20 \, \mu m$ and an $6^{th}$ order super-Gaussian pulse for the calculations. Fig.2 (b) displays the evolution of populations $\rho_{11}, \rho_{22}$ and $\rho_{33}$ with the nonlinear positive chirp rate. We have assumed that all the atoms are initially in the state $|1\rangle$. Such initial preparation of the system can easily be achieved in case of numerous physical systems by the means of optical pumping using a resonant CW laser. It is evident that one can achieve a near complete transfer of population (99.74 %) from state $|1\rangle$ to state $|2\rangle$ with the few-cycle pulse shown in Fig.2 (a).

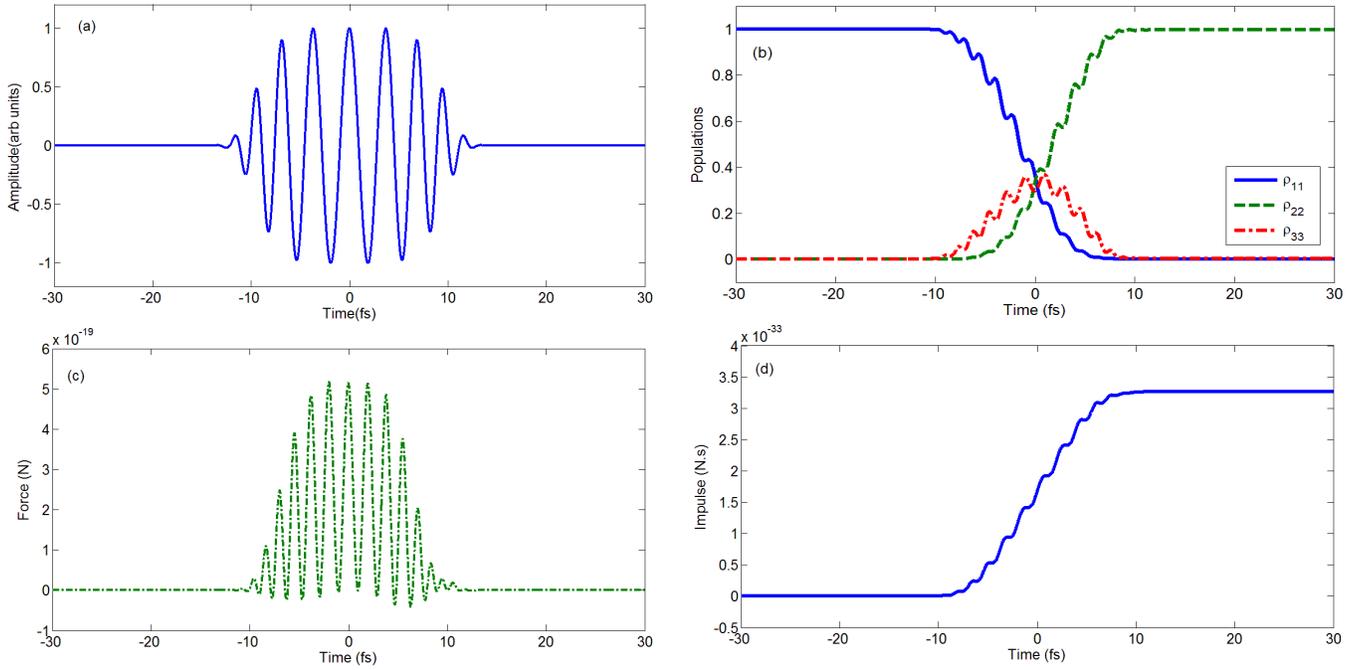

FIG.2. (Color online) (a) Pulse profile with $\chi = +0.0037\ fs^{-3}$ (b) Temporal evolution of populations (c) Optical dipole force experienced by the atoms (d) Total momentum imparted to the atoms as a function of time.

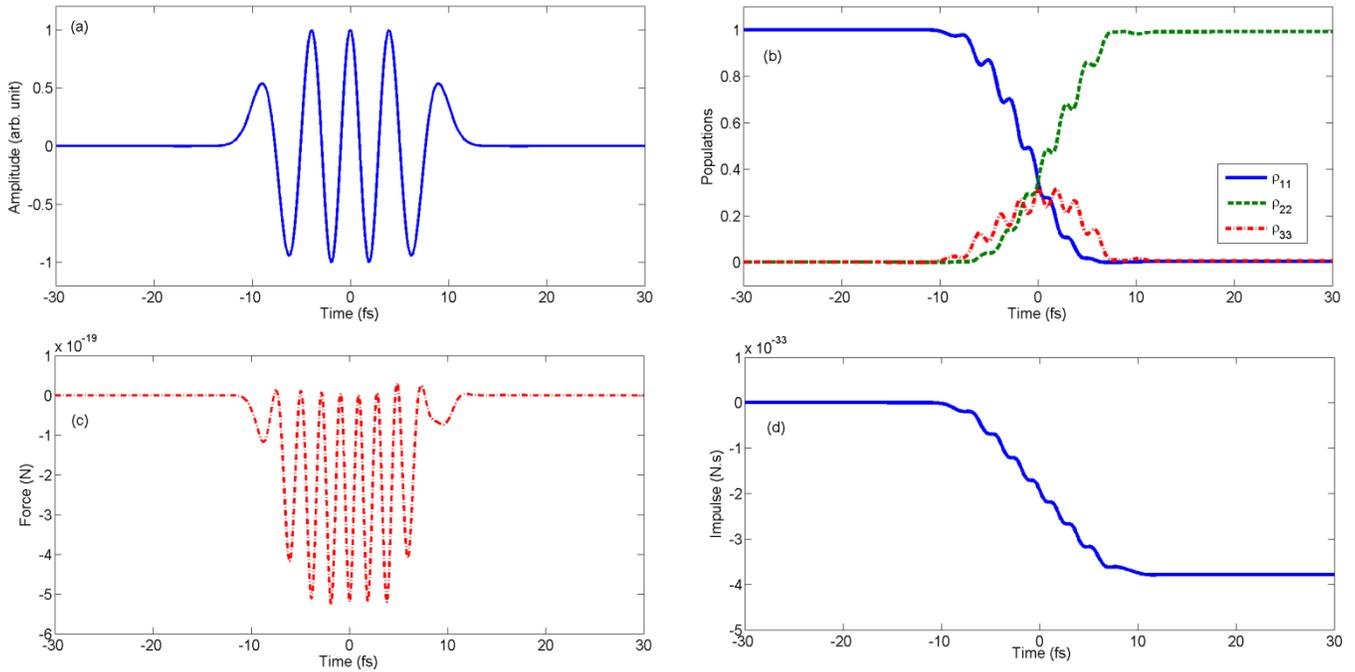

FIG.3. (Color online) (a) Pulse profile with $\chi = -0.0037\ fs^{-3}$ (b) Temporal evolution of populations (c) Optical dipole force experienced by the atoms (d) Total momentum imparted to the atoms as a function of time.

As the pulse interacts with the atomic system, atoms experience a time dependent optical dipole force due to the non-zero gradient of the field amplitude. It is clear from the force-time plot of Fig. 2 (c) that even though oscillatory in nature, the force remains positive during the entire pulse duration and thus a net positive linear impulse is given to the atoms as shown in Fig. 2 (d). This force can thus be used to expand an atomic beam. Fig. 3 (a) - (d) shows the corresponding plots for the negatively chirped pulse. Here it is interesting to note that while a near complete transfer of population (99.1 %) is achievable, the force acting on the atoms is negative and hence can be used to focus an atomic beam. It is important from the application point of view to consider the effect of the pulse over its entire cross-sectional area. In passing, it may be useful to briefly discuss the physics of population transfer with nonlinearly chirped pulses. The mechanism of population transfer with the considered nonlinearly chirped pulses is quite different compared to the linearly chirped pulses. In the proposed work, the population transfer is non-adiabatic in contrast to the adiabatic population transfer with the so called adiabatic rapid passage (ARP) and stimulated Raman adiabatic passage (STIRAP) schemes. In our study, the adiabatic condition $\Omega\tau \gg 2\pi$ is not fulfilled owing to the chosen pulses parameters. The consequence of non-adiabatic evolution of population dynamics is that the population in quantum state $|3\rangle$ during the intermediate time is approaching a large value as the adiabatic criteria is not fulfilled for the chosen laser pulse areas. However, finally the quantum state $|2\rangle$ receives almost all the populations as can be seen from Figs. 2(b) and 3(b). The chosen nonlinear chirp provides the robustness in the population transfer against the variation of pulse parameters [42]. The population dynamics may depend on the initial state preparation. In the proposed study, however, we assume that initially all the populations reside in ground state $|1\rangle$ which is practically possible. The proposed scheme for the coherent population transfer is quite similar to the stimulation emission pumping (SEP) scheme. In SEP scheme [32], the CW lasers interact simultaneously with both the transition paths in $\Lambda-$like atomic systems. However, the maximum amount of population transfer with SEP scheme is limited due to the presence of relaxation processes during the population transfer. For example, one can achieve maximum 30 % population transfer in $\Lambda-$like three level atomic systems with SEP technique [32]. On the other hand, in our scheme, we have shown nearly complete population transfer in the time scale where relaxation processes are negligible. With the pulse parameters same as in Fig. 2(a) we investigate the final population of the state |2>, $\rho_{22}(\infty)$ and the total final impulse imparted to the atoms as a function of their spatial location on the transverse plane. The results are shown in Fig. 4(a) and Fig 4(b) respectively. A very high population transfer is achieved up to distances of about 70 $\mu m$ from the axis while a near constant total impulse is imparted to the atoms up to distances of about 50 $\mu m$ from the centre of the pulse. It is worthwhile to note here that the optical force on the atoms much farther away from the beam axis is about three orders of magnitude greater than that near the axis. Also the population transfer from |1> to |2> is incomplete farther away from the axis. Hence the proposed scheme can in fact be used to select the part of the atomic beam with a very high population in the state |2>.

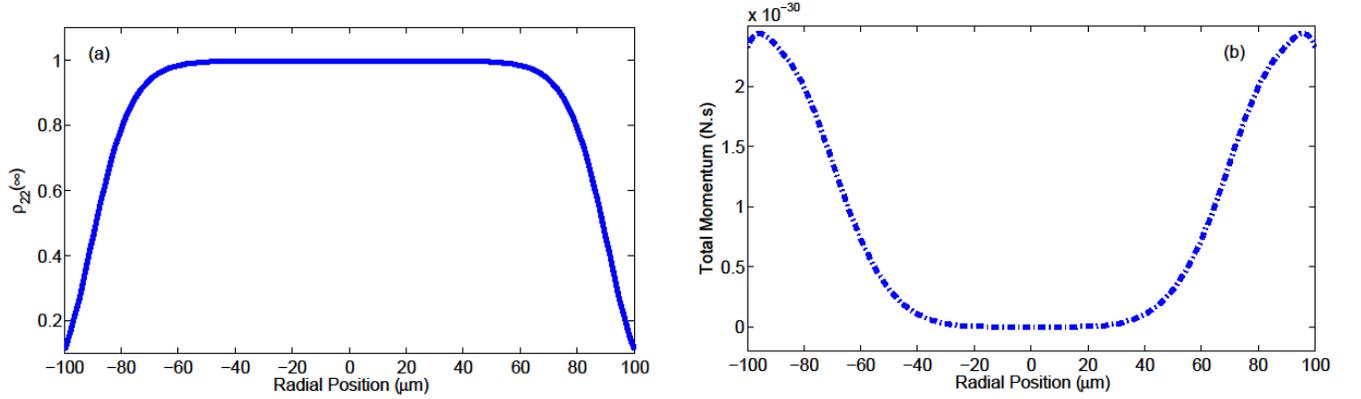

FIG.4. (Color online) (a) Final population of the quantum state |2> as a function of the spatial location ($r$) on the transverse plane (b) Total momentum imparted to the atoms at the spatial location ($r$).

We demonstrate in Fig. 5(a) that the final population transfer to the quantum state |2>, $\rho_{22}(\infty)$ is robust against the variation in Rabi frequency ($\Omega$) and pulse duration($\tau_0$) of the few-cycle pulse. Fig. 5(b) displays the robustness of the proposed scheme against variations in the chirp rate ($\chi$) and Rabi frequency($\Omega$). It is important to note here that the final population of the state |2> does not depend critically on the relative values of the two Rabi frequencies $\Omega_{13}$ and $\Omega_{23}$.

We have shown the robustness of this scheme against the variations in the two Rabi frequencies, with other parameters of the pulse same as in Fig.2, in Fig. 5 (c). In fact with a judicious choice of other pulse parameters like the chirp rate, the pulse duration, central frequency and the order of the pulse profile, one should be able to implement the proposed scheme in a given atomic system.

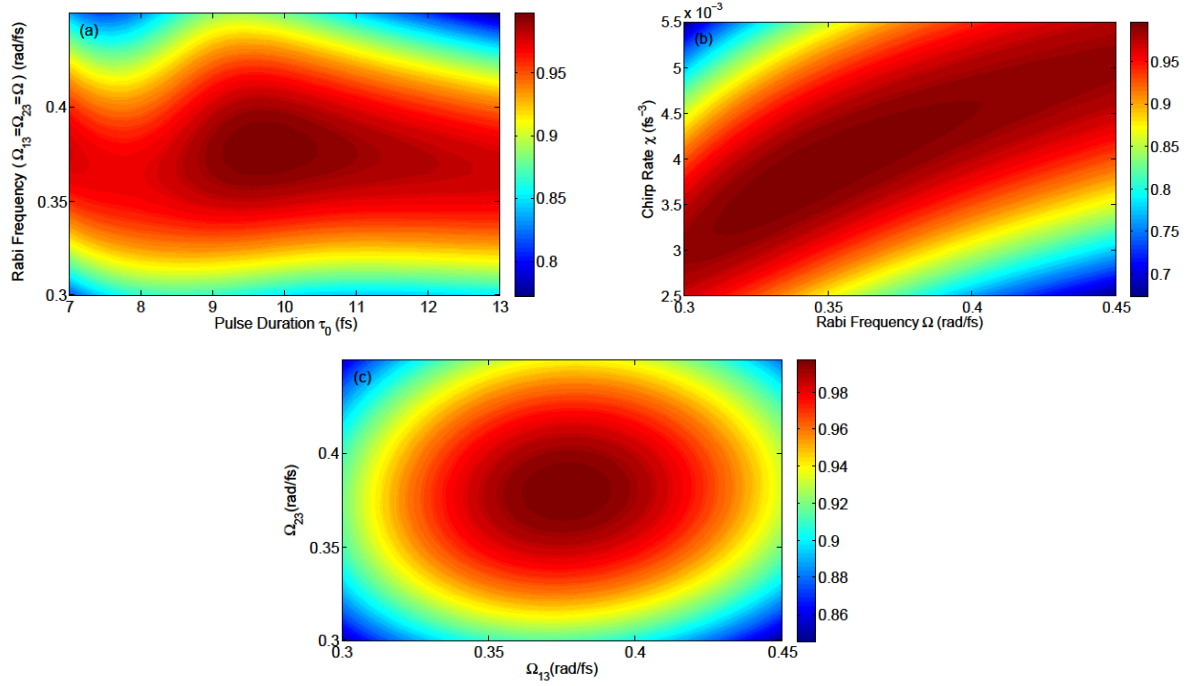

FIG.5. (Color online) Final population transfer $\rho_{22}(\infty)$ to the quantum state |2> as a function of (a) Pulse duration $\tau_0$ and Rabi Frequency $\Omega$ (b) Rabi frequency $\Omega$ and Chirp rate $\chi$ (c) the two Rabi frequencies, $\Omega_{13}$ and $\Omega_{23}$.

It is instructive to examine the effect of the order of the super-Gaussian shaped few-cycle femtosecond pulses on the transfer of population to the state |2>. As is clear from the Fig. 6(a), with higher order pulses one can achieve better control over the population transfer process over a large part of the cross sectional plane of the pulse. The other parameters used in the simulation are the same as that in Fig.2. Fig. 6(b) displays the corresponding plot of the total momentum received by the atoms due to the optical dipole force of the chirped pulse interacting with it. Again it is evident that with the pulses of higher order it is possible to better select the portion of the atomic beam where the population transfer is nearly complete.

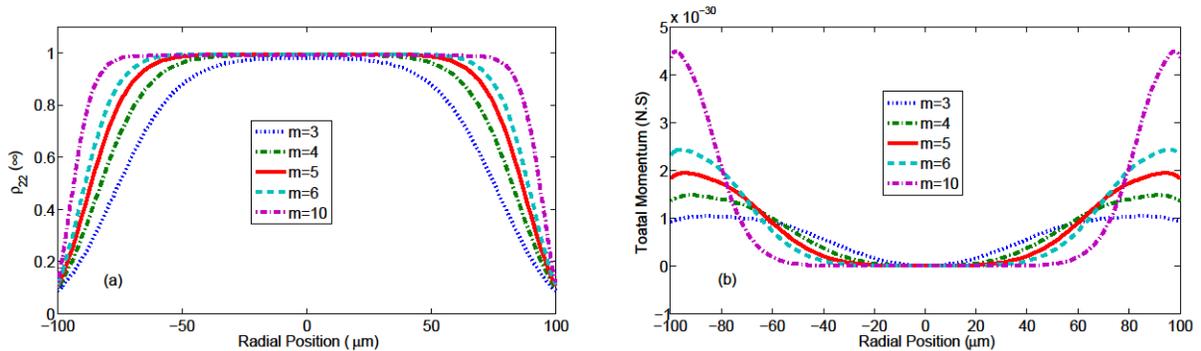

FIG.6 (a) (Color online) Final Population of the state |2> as a function of the spatial location (r) on the cross sectional plane of the pulses with different order profiles (b) Total linear momentum gained by the atoms positioned at location (r) on the cross sectional plane. All other parameters same as that used in Fig.2.

We now report the results obtained by treating the $\Lambda$–system with a series of few-cycle pulses. The atomic parameters used here are the same as in the single pulse case reported above. We use pulses with super-Gaussian profile of order 6 in all these calculation. The pulse repetition period is taken to be $T = 2\pi n/\omega_{21} \approx 10\ ns$, where $n$ is an integer. The chirp rate, $\chi = -0.002\ fs^{-3}$, and the Rabi frequency is $\Omega = 0.37\ rad/fs$ for the individual pulses in the pulse train. All other pulse parameters are same as in the single pulse case presented above. The point under investigation is assumed to be located at $r = 70\ \mu m$ from the beam axis. Initially all the atoms are assumed to be in the state $|1>$. In Fig.7 (a) we display the time evolution of the coherence between the states $|1>$ and $|2>$. The system starts out with zero coherence, but with each passing pulse coherence builds up in the system and $\rho_{12}$ converges to -0.5, i.e. attains maximum coherence, after interacting with about 130 pulses. As is clear from the figure, the imaginary part of the coherence fluctuates initially but dies out fast. We show the population transfer from the bright state to the dark state involved in this process in Fig.7 (b). The system starts with equal population in the bright and the dark state but with passage of each pulse, population from the bright state accumulates in the dark state and once the population in the bright state reaches zero the medium does not interact with the radiation any more.

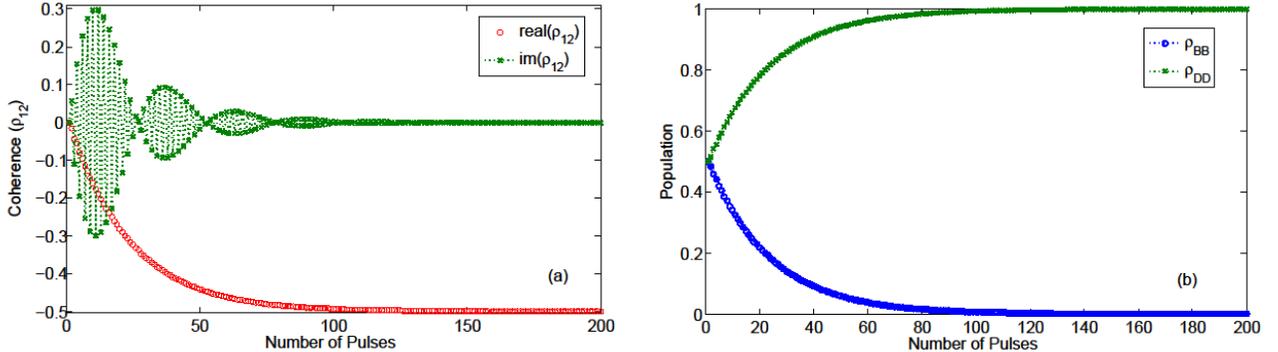

FIG.7 (Color online) (a) Accumulation of coherence between state $|1>$ and $|2>$ by a train of femto-second pulses. (b) Transfer of population from the bright state to the dark state leading to transparency of the medium to the radiation.

The effect of the chirp rate of the pulses on accumulation of coherence $\rho_{12}$ in the system and optical dipole force experienced by the atoms is investigated for both the negative and positive nonlinear chirping cases. The dependence of the coherence on the chirp rate and the number of pulses is given in Fig 8. All relevant parameters other than chirp rate were taken to be the same as in Fig 7.

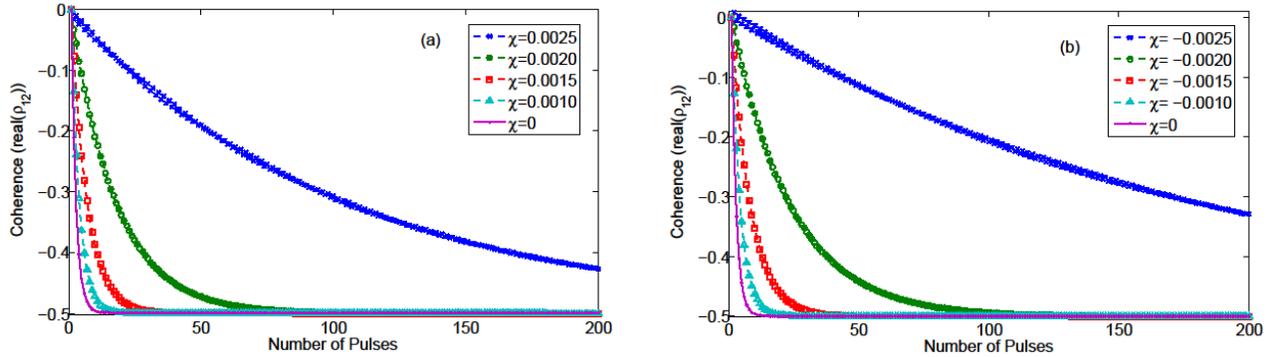

FIG 8. (Color online) Accumulation of coherence in the Λ-system with a series of pulses (a) Positive Chirp rate (b) Negative Chirp rates ($in\ fs^{-3}$)

We now show that in addition to creating coherence between the states |1> and |2> in the Λ−system, it is possible to control the trajectory of the atoms in the atomic beam. The trajectory of the atoms, interacting with the pulse train, on the transverse cross section is shown in Fig 9. We can see that the beam gets focused and defocused by interaction with the pulse train with negative and positive chirping rates respectively. The mass of the atom was assumed to be $m = 23\ a.m.u$ for the trajectory calculation presented here.

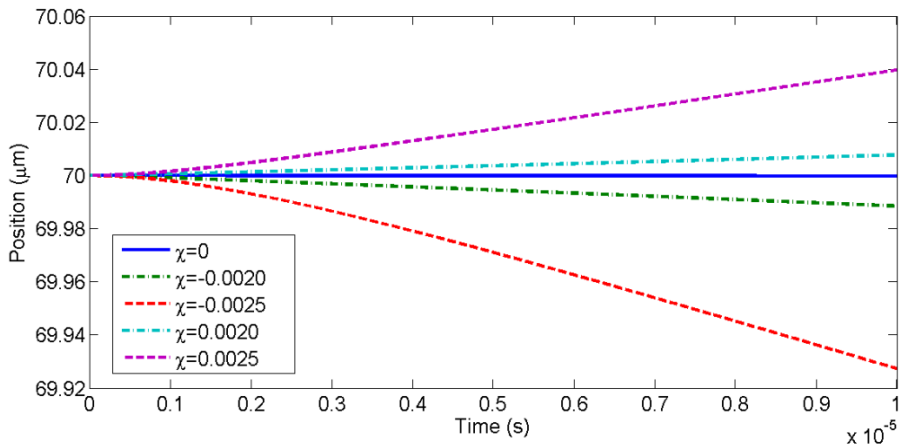

FIG.9. (Color online) Motion of the atoms under the influence of the optical dipole force on the transverse plane of the atomic beam

It is of worth to note here that while one can achieve maximum coherence by using non-chirped pulses in the pulse-train with a relatively few number of pulses, the amount of force experienced by the atoms and hence the deflection is negligible. On the other hand, it takes larger number of pulses to establish maximum coherence if one uses greater chirp rates in case of both positive and negative nonlinear chirping. However the degree of focus/defocus in the latter case is appreciable and controllable. The scheme may be applicable to some alkali atoms, atomic systems like Boron, ionized neon and molecular systems, subject to judicious choice of parameters.

## IV. CONCLUSIONS

We have studied optical dipole force in a $\Lambda$–like three level atomic system using a linearly chirped super-Gaussian femtosecond pulse. We show that the same scheme could be used to obtain almost complete population transfer to the state |2> of the atom. Also, using a train of femtosecond pulses we show that in addition to creating maximum coherence between the ground state |1> and the excited state |2>, it is possible to control the trajectory of the atoms in an atomic beam. The scheme is found to be robust against the variations in the Rabi frequencies and pulse parameters like the chirp rate and the pulse duration.

## ACKNOWLEDGMENTS

A.K.S. would like to acknowledge the financial support from CSIR (Grant No. 03(1252)/12/EMR-II).